# Characterization of large area photomultiplier ETL 9357FLB for liquid argon detector

DU Ying-Shuai(杜迎帅)[1], YUE Qian(岳骞)[2;3], LIU Yi-Bao(刘义保)[1], CHEN Qing-Hao(陈庆豪)[2;3], LI Jin(李金)[2;3], CHENG Jian-Ping(程建平)[2;3], KANG Ke-Jun(康克军)[2;3], LI Yuan-Jing(李元景)[2;3], LI Yu-Lan(李玉兰)[2;3], MA Hao(马豪)[2;3], XING Hao-Yang(幸浩洋)[4], YU Xun-Zhen(余训臻)[4], ZENG Zhi(曾志)[2;3]

[1] Engineering Research Center of Nuclear Technology Application, Ministry of Education, East China Institute of Technology, Nanchang, 330013,China

[2] Department of Engineering Physics, Tsinghua University, Beijing 100084, China

[3] Key Laboratory of Particle and Radiation Imaging, Tsinghua University, Ministry of Education, China

[4] School of Physical Science and Technology, Sichuan University, Chengdu, 610041, China

**Abstract:** The China Dark Matter Experiment (CDEX) Collaboration will carry out a direct search for weakly interacting massive particles with germanium detectors. Liquid argon will be utilized as an anti-Compton and cooling material for the germanium detectors. A low-background and large-area photomultiplier tube (PMT) immersed in liquid argon will be used to read out the light signal from the argon. In this paper we carry out a careful evaluation on the performance of the PMT operating at both room and cryogenic temperatures. Based on the single photoelectron response model, the absolute gain and resolution of the PMT were measured. This has laid a foundation for PMT selection, calibration and signal analysis in the forthcoming CDEX experiments.

**Keyword:** Dark matter, PMT, SER, Gain, Dark count
**Pacs:** 85.60.Ha, 29.40.Mc, 07.20.Mc

## 1 Introduction

The China Dark matter experiment (CDEX) is going to directly detect the flux of weakly interacting massive particles (WIMPs) with a point-contact germanium detector array, where liquid argon is used as the cooling system, as well as both passive and active shielding[1,2,3]. PMTs of the type ETL 9357FLB are directly immersed in liquid argon to read out its scintillation light signals. It is well known that the ambient temperature strongly influences the performances of the PMT, So it is necessary to test and study the performance of the PMT at liquid argon cryogenic temperature.

For this work, we built an experiment setup, and studied the performance of the PMT using the single photoelectron response (SER) model. Instead of liquid argon, liquid nitrogen was used to do the tests, as its temperature is close to that of liquid argon and the price is cheaper. Measurements were carried out at both room (289.7K) and cryogenic (77K) temperatures. We obtained the absolute gain, energy resolution, and dark rate of the PMT, and their changes with temperature.

## 2 Experimental setup

The experimental setup is sketched in Fig.1. One PMT was housed inside a dewar which can be filled with liquid or gaseous nitrogen. A blue light signal from the external LED is guided to the PMT by means of an optical fiber. The LED, whose central emission wavelength is 420nm, was driven by a pulse signal generated by an arbitrary generator (AFG). The output end from the optical fiber was aligned to the center of the PMT tube. The frequency of the pulse light beam was adjusted from the AFG and the number of photons within each beam was controlled by tuning the amplitude of each input pulse.

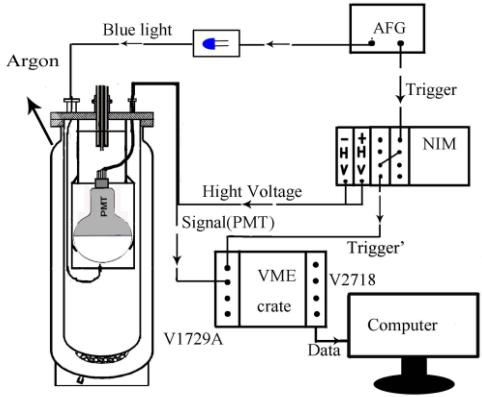

Fig1 Sketch of the experimental setup.

The high voltage (HV) of the PMT was supplied on the cathode and dynode chain through a voltage divider circuit (Fig.2) to distribute the proper electrical potential suggested by the manufacturer. We applied the voltage distribution as follows: the first dynode was connected to ground; the cathode was fixed to -600V; and the positive voltage on the dynode chain from D1 to D12 went from 600V to 1100V. The two HV power supplies, one for the cathode and the other for the dynode-chain, were used independently. The signal output was read out directly through an AC-coupled capacitor (C5, 10nf), and the pulse data was recorded by a flash analog-to-digital converter (CAEN V1729A). The VME master module V2718 then transferred the data to the computer through optical fiber.

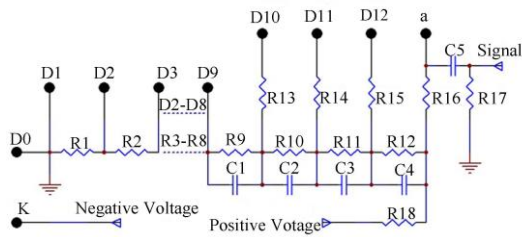

Fig2 Voltage divider circuit of the PMT.

## 3 Experiments and Results

### 3.1 Single photoelectron response (SER)

The SER, i.e., the charge distribution of the PMT pulses integrated over the whole signal pulse shape, is directly related to the PMT multiplication processes and can be affected by the operating temperature [4]. The number of photoelectrons (p.e.) induced by incident scintillation light and emitted from the PMT cathode follows a Poisson distribution. If the possibility of double photoelectron signals can be controlled less than, for instance, 10% of single photoelectron signals, then most of the photoelectron signals are single photoelectron signals. We call this the single photoelectron level, and the level can be achieved by adjusting the intensity of the blue light signals [5]. When the dynode multiplication factor is large enough (above 4), and the efficiency of the secondary electron collection is close to 1, the PMT response to photoelectrons can be treated as a Gaussian function. Assuming that the PMT response is linear, we can then get the ideal function of the PMT response to low intensity light sources as follows, with the exponential part being background noise and the Gaussian parts being the sum of the SER (n=1) and multi-p.e. responses (n$>$1)[6]:

$$S_{Ideal}(x) = (1-\omega)\sum_{n=0}^{\infty}\frac{\mu^n e^{-\mu}}{n!}\frac{1}{\sigma_1\sqrt{2n\pi}}\exp\left(-\frac{(x-n\mu_1)^2}{2n\sigma_1^2}\right) + \omega\alpha e^{-\alpha x} \quad (1)$$

where $\omega$ is the amplitude of the exponential part; $\alpha$ is the exponential decay constant; n is the number of the photoelectrons; $\mu$ is the mean value of the number of photoelectrons and $\mu_1$ and $\sigma_1$ are the mean value and the standard deviation of the Gaussian part of the SER, respectively.

The positive pulse signal of the LED driver is set to a width of 10ns and amplitude of 1.78V in our experiment. The LED was set to emit at a certain frequency. Each signal of the PMT output was collected and analysed, allowing us to obtain the charge spectrum of the PMT response to a low intensity light source. An example of a typical SER charge spectrum is shown in Fig.3. The PMT response distribution had been fitted with Equation (1), which contains an exponential and two Gaussian functions: the first Gaussian takes into account the response of the PMT to a single photoelectrons, and the second accounts for event with 2 photoelectrons. $\mu_1$ is the size of the charge of the PMT response to single photoelectrons, and is also the absolute gain of the PMT for single photoelectrons. The resolution is obtained from $\sigma_1/\mu_1$ and $V_{peak}/V_{valley}$, where $V_{peak}$ and $V_{valley}$ are the value of the peak and valley shown in Fig.3.

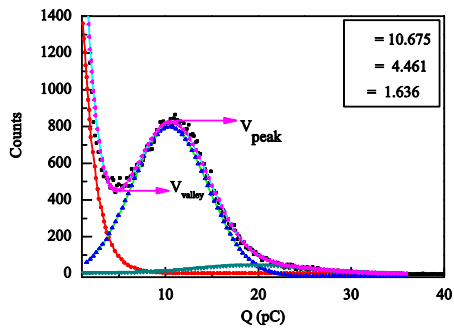

Fig3 Example of a SER spectrum as obtained for the ETL 9357FLB at 1500 V at Liquid nitrogen temperature.

### 3.2 Gain

The gain is obtained from the position of the single photoelectron peak in the charge spectrum. We studied the PMT gains for different dynode-chain voltages and at different temperatures. The results obtained from the same PMT and a linear fit to the data are plotted in Fig.4. The signal gain is represented by an exponential behavior as a function of high voltage for both operating temperatures. The reduction in the gain at liquid nitrogen temperature is appreciable, as shown in Fig.4, and there is good linearity between gain and voltage in the range 1300V-1600V, with the R-Square values of 0.997 at room temperature and 0.997 at cryogenic temperature.

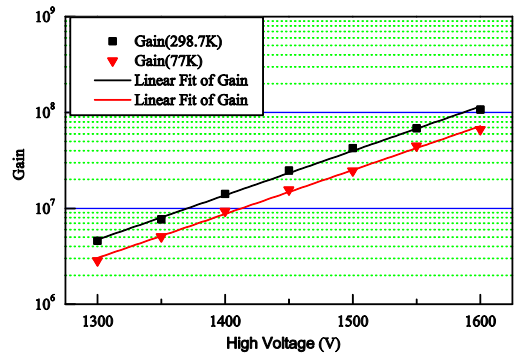

Fig4 Signal gain as a function of voltage at room and Liquid nitrogen temperatures.

### 3.3 SER Resolution and Peak-to-Valley Ratio

The ratio of $\sigma_1$ to $\mu_1$, which is the SER resolution of the PMT, can be obtained from the fitting results and is another important characteristic to be measured. Fig.5 presents the resultion obtained as a function of the gain at different temperatures. The SER resolution is found to be 0.43, and changes only slightly in the range of gains we tested.

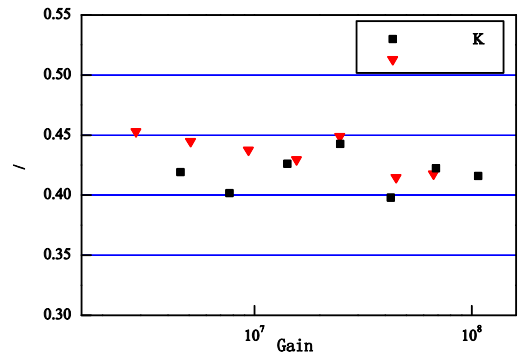

Fig5 Relationship between $\sigma_1/\mu_1$ and gain at room and liquid cryogenic temperatures.

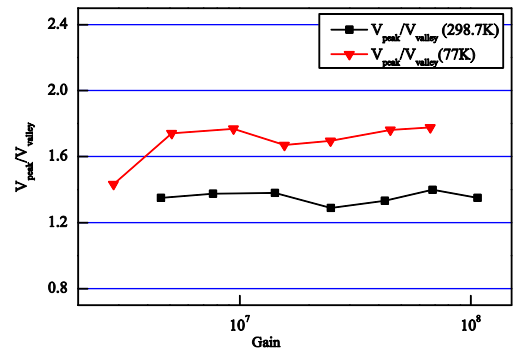

Fig6 $V_{peak}/V_{valley}$ vs gain at room and liquid nitrogen temperature.

The ratio of $V_{peak}/V_{valley}$, obtained from the charge spectrum, is plotted in Fig.6. As shown in the plot, the value of $V_{peak}$ to $V_{valley}$ at cryogenic temperature is higher than that at room temperature because the noise at cryogenic temperature is lower. The ratio changed little over the range we measured, because the PMT amplified not only the light signal but also the background noise. We should therefore reduce the background noise to increase the value of $V_{peak}$ to $V_{valley}$.

### 3.4 Dark Count Rate

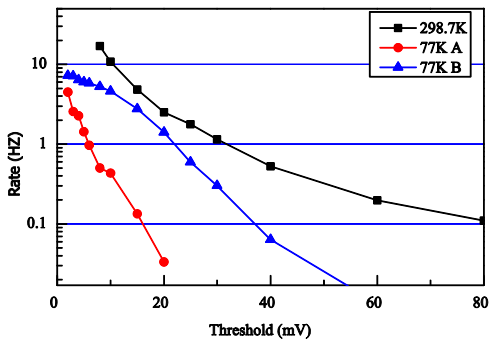

Fig.7 Plot of dark counts as a function of threshold at room and cryogenic temperature. "298.7K" indicates the dark counts at room temperature, "77K A" indicates the dark counts for the PMT wrapped with black cloth and immersed in nitrogen, multiplied by 10, and "77K B" indicates the dark counts for the PMT immersed in nitrogen directly.

Measurements of PMT dark counts were carried out at room and cryogenic temperatures. The output pulses from the PMT under test, operating in conditions of complete darkness, were discriminated and counted during repetitive cycles in which the discrimination threshold was adjusted from 1 to 200mV and the high voltage was adjusted from 1200 to 1800V. We found our results were different from Ref.[7], in which the dark count rate of the PMT in liquid nitrogen increased. We therefore performed two experiments at cryogenic temperature, first with the PMT immersed directly in the liquid nitrogen and second with the PMT immersed in liquid nitrogen while wrapped in black cloth, thus cutting off light coming from the liquid nitrogen or from impurities. The rate of dark counts at different thresholds is been plotted in Fig.7, with a high voltage of 1400V.

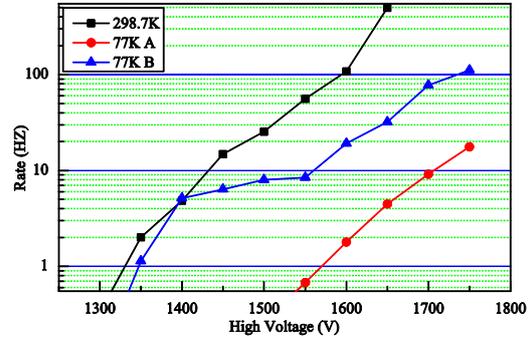

Fig8 Plot of dark counts as a function of high voltage at room and cryogenic temperature.

The rate of dark counts for different high voltages is shown in Fig.8. The threshold was set at 15mV. From all of these results we can conclude that the dark counts decrease at cryogenic temperatures, but some light produced in the liquid nitrogen or its impurities may increase the dark counts.

## 4 Conclusions

In this paper we have discussed the SER model and analyzed the performance of the ETL 9357FLB PMT at room and cryogenic temperatures. The results obtained demonstrate that: (1) PMT absolute gain can reach $10^7$ at both room and cryogenic temperature, but the cryogenic environment will reduce the gain of the PMT slightly; (2) the energy resolution is 40%-45%, with very slight variation, at both room and cryogenic temperature, without much difference between them; (3) the rate of $V_{peak}$ to $V_{valley}$ at room temperature is lower than that at cryogenic temperature; (4) dark counts will be lower at cryogenic temperature, and any light coming from the liquid nitrogen or its impurities will influence PMT dark counts. Overall, we can conclude that the PMT gain drops at cryogenic temperature but the resolution is better and dark counts are lower. This experiment has studied the performances of the photomultiplier tubes, and provided an effective solution for testing the batch of photomultiplier tubes for the CEDX10 experiment.